\documentclass[aps,pre,twocolumn,superscriptaddress]{revtex4-2}

\usepackage{amsmath,amssymb}
\usepackage{graphicx}
\usepackage{hyperref}
\usepackage{booktabs}
\usepackage{xcolor}
\usepackage{multirow}

\newcommand{\Tc}{T_{\mathrm{c}}}
\newcommand{\ftopo}{f_{\mathrm{topo}}}
\newcommand{\TP}{\mathrm{TP}}
\newcommand{\PE}{\mathrm{PE}}

\begin{document}

\title{How much of persistent homology is topology?\\A quantitative decomposition for spin model phase transitions}

\author{Matthew Loftus}
\affiliation{Cedar Loop LLC, Washington, USA}

\date{\today}

\begin{abstract}
Point-cloud persistent homology (PH)---computing alpha or Rips complexes on spin-position point clouds---has been widely applied to detect phase transitions in classical spin models since Donato \textit{et al.}~(2016), with subsequent studies attributing the detection to the topological content of the persistence diagram. We ask a simple question that has not been posed: what fraction of the PH signal is genuinely topological? We introduce $\ftopo$, a quantitative decomposition that separates the density-driven and topological contributions to any PH statistic by comparing real spin configurations against density-matched shuffled null models. Across the 2D Ising model (system sizes $L = 16$--$128$, ten temperatures) and Potts models ($q = 3, 5$), we find that $H_0$ statistics---total persistence, persistence entropy, feature count---are 94--100\% density-driven ($\ftopo < 0.07$). The density-matched shuffled null detects $\Tc$ at the identical location and with comparable peak height as real configurations, showing that density alone is sufficient for phase transition detection. However, $H_1$ statistics are partially topological: the topological fraction grows with system size as $\delta(\TP_{H_1}) \sim L^{0.53}$ and follows a finite-size scaling collapse $\delta(T, L) = L^{0.53}\, g(tL^{1/\nu})$ with collapse quality CV $= 0.27$. The longest persistence bar is strongly topological ($\ftopo > 1$) and scales with the correlation length. A scale-resolved analysis reveals that the topological excess shifts from large-scale to small-scale features as $L$ increases. We propose that the TDA-for-phase-transitions community adopt shuffled null models as standard practice, and that $H_1$ rather than $H_0$ statistics be used when genuine topological information is sought.
\end{abstract}

\maketitle

\section{Introduction}
\label{sec:intro}

Persistent homology (PH) has become a standard tool for detecting phase transitions in classical spin models~\cite{donato2016,cole2021,sale2022,hirakida2021,olsthoorn2020,spitz2020}. A prevalent methodology constructs point clouds from spin positions and computes the persistence diagram (PD) of an alpha or Rips complex filtration, showing that a PD-derived statistic exhibits a peak or transition at the critical temperature $\Tc$. This point-cloud approach has been applied to Ising~\cite{donato2016}, Potts and XY models~\cite{sale2022}, and lattice gauge theories~\cite{hirakida2021,spitz2020}. (A separate branch of the literature uses sublevel-set filtrations on the spin field directly~\cite{cole2021,olsthoorn2020}; our analysis addresses only the point-cloud methodology, which dominates the literature.)

Despite the success of this program, a fundamental question has not been asked: \textit{how much of the PH signal is actually topological?} When PH is applied to the point cloud of majority-spin sites, the number and density of points varies with temperature---the magnetization, and thus the point count, changes across the transition. Any PH statistic that depends on the number of points will therefore change across the transition, regardless of whether the \textit{spatial arrangement} of those points carries topological information.

In this paper, we introduce a quantitative framework to answer this question. For each spin configuration, we construct a \textit{density-matched shuffled null}: random lattice sites with the same point count as the majority-spin cloud, but no spatial correlations. Any PH statistic that matches between the real configuration and its shuffled null is density-driven; any discrepancy is topological. We define the topological fraction
\begin{equation}
\ftopo = \frac{S_{\mathrm{real}} - S_{\mathrm{shuf}}}{S_{\mathrm{real}} - S_{\mathrm{rand}}},
\label{eq:ftopo}
\end{equation}
where $S_{\mathrm{rand}}$ is the statistic evaluated on a random point cloud at reference density $\rho = 1/2$ (Ising) or $\rho = 1/q$ (Potts).

Our main findings are:
\begin{enumerate}
\item $H_0$ statistics (components) are 94--100\% density-driven: $\ftopo(\TP_{H_0}) < 0.07$, $\ftopo(\PE) \approx 0.05$, $\ftopo(n_{H_0}) = 0.000$ at all system sizes.
\item The shuffled null detects $\Tc$ at the identical temperature as the real configurations, demonstrating that density alone is sufficient for phase transition detection.
\item $H_1$ statistics (loops) are partially topological, and the topological fraction \textit{grows} with system size: the relative topological excess scales as $\delta(\TP_{H_1}) \sim L^{0.53}$ and satisfies a finite-size scaling collapse.
\item The maximum persistence bar is strongly topological and scales with the correlation length $\xi$.
\item The pattern holds across three models: 2D Ising and Potts $q = 3, 5$.
\end{enumerate}

These results synthesize evidence from six experiments performed over six months of systematic null-model testing. We show that the density confound is not an artifact of a particular implementation but a fundamental property of point-cloud PH applied to spin systems. Our framework provides the TDA-for-phase-transitions community with a standardized null model and identifies what PH genuinely contributes: $H_1$ loop structure and maximum persistence, not the bulk statistics commonly reported.

\section{Methods}
\label{sec:methods}

\subsection{Spin models and sampling}
\label{sec:models}

We study the 2D Ising model on square lattices of linear size $L \in \{16, 24, 32, 48, 64, 128\}$ at ten temperatures spanning the ordered ($T < \Tc$) and disordered ($T > \Tc$) phases, where $\Tc = 2/\ln(1 + \sqrt{2}) \approx 2.269$ is the exact Onsager value~\cite{onsager1944}. We also study the 2D Potts model with $q = 3$ (second-order transition, $\Tc \approx 0.995$) and $q = 5$ (first-order transition, $\Tc \approx 0.852$) at $L = 48$, each at ten temperatures. Configurations are sampled using the Wolff cluster algorithm~\cite{wolff1989}, with cold-start thermalization for $T \leq \Tc$. Between measurements, we perform 5--10 decorrelation sweeps depending on system size.

\subsection{Point clouds and persistence diagrams}
\label{sec:ph}

For each spin configuration, we identify the majority-spin species (the spin value with the largest count) and extract the lattice positions of all sites carrying that value. This produces a point cloud of $n = \rho L^2$ points in $[0, L]^2$, where $\rho$ is the majority fraction. The alpha complex filtration~\cite{edelsbrunner1994} is computed using GUDHI~\cite{gudhi2015}, yielding a persistence diagram with $H_0$ (connected component) and $H_1$ (loop) bars.

From each diagram we extract: total persistence $\TP_{H_k} = \sum_i (d_i - b_i)$ for $k = 0, 1$; persistence entropy $\PE = -\sum_i p_i \ln p_i$ where $p_i = (d_i - b_i) / \TP$; feature counts $n_{H_k}$; maximum bar length; and mean bar length.

\subsection{Null models}
\label{sec:null}

For each real configuration with $n$ majority-spin sites, we construct two null models:

\textit{Shuffled null} (density-matched): select $n$ lattice sites uniformly at random from all $L^2$ sites. This preserves the point count (and thus the density $\rho$) but destroys all spatial correlations.

\textit{Random null} (fixed baseline): select $L^2/2$ lattice sites uniformly at random (Ising) or $L^2/q$ sites (Potts), providing a temperature-independent reference at the disordered-phase density.

Both null models are on the same lattice, ensuring identical geometric substrate. The shuffled null is the key comparison: any PH statistic that matches between the real and shuffled configurations is fully determined by the density $\rho$, not by spatial structure.

\subsection{The topological fraction $\ftopo$}
\label{sec:ftopo}

We define $\ftopo$ as in Eq.~\eqref{eq:ftopo}. When $\ftopo \approx 0$, the statistic is density-driven (shuffled matches real). When $\ftopo \approx 1$, it is topological (shuffled matches random, real is different). Values $\ftopo > 1$ indicate the shuffled null undershoots the random baseline, which occurs when higher density compresses bar lengths below the $\rho = 1/2$ reference.

We compute $\ftopo$ from ensemble means over 100--200 configurations per $(L, T)$ pair, with bootstrap standard errors (1000 resamples). For diagnosing the mechanism, we also use the relative topological excess
\begin{equation}
\delta = \frac{S_{\mathrm{real}} - S_{\mathrm{shuf}}}{S_{\mathrm{shuf}}},
\label{eq:delta}
\end{equation}
which avoids the random-baseline denominator and is better suited for finite-size scaling analysis.

\section{Results}
\label{sec:results}

\subsection{$H_0$ is density-driven}
\label{sec:h0}

Table~\ref{tab:ftopo} presents $\ftopo$ for all seven PH statistics at $\Tc$. The key finding is that $H_0$ statistics are overwhelmingly density-driven:

\begin{itemize}
\item $\ftopo(\TP_{H_0}) = 0.037 \pm 0.004$ at $L = 64$ and $0.065 \pm 0.007$ at $L = 128$: density accounts for 93--96\% of $H_0$ total persistence.
\item $\ftopo(\PE) \approx 0.05$ at all $L$, with no significant size dependence.
\item $\ftopo(n_{H_0}) = 0.000$ at all $L$: the $H_0$ feature count is \textit{exactly} determined by the point count $n$, because the alpha complex on $n$ generic 2D points always produces exactly $n - 1$ finite $H_0$ bars~\cite{edelsbrunner2010}.
\end{itemize}

The real-shuffled gap for $\TP_{H_0}$ is statistically marginal (1.1$\sigma$ at $L = 64$, 2.1$\sigma$ at $L = 128$), consistent with a small but growing topological correction. In contrast, the real-shuffled gap for the maximum bar length exceeds 17$\sigma$ at $L = 64$ and 26$\sigma$ at $L = 128$, confirming that spatial correlations strongly affect the longest-lived feature but not the bulk statistics.

\begin{figure}[t]
\includegraphics[width=\columnwidth]{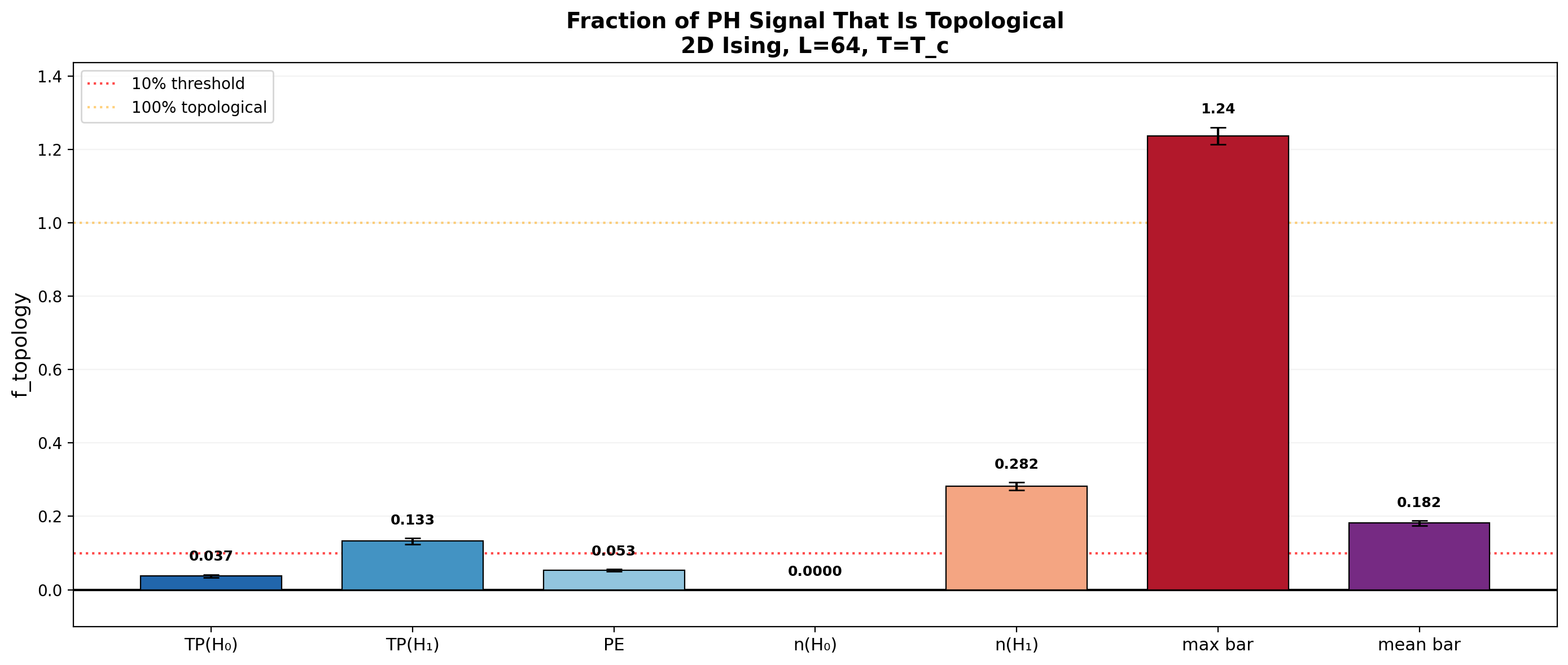}
\caption{Topological fraction $\ftopo$ for each PH statistic at $\Tc$, $L = 64$. Blue bars: standard PH statistics. Red/green/yellow: per-bar split. $H_0$ statistics (TP, PE, $n$) have $\ftopo < 0.06$. The maximum bar has $\ftopo > 1$ (spatial correlations enhance it beyond the random baseline). $H_1$ count has $\ftopo = 0.28$.}
\label{fig:money}
\end{figure}

\begin{table}[t]
\caption{Topological fraction $\ftopo$ at $\Tc$ for the 2D Ising model. Errors (not shown for readability) are $\pm$0.003--0.010 from bootstrap. The max bar row uses the relative excess $\delta$ [Eq.~\eqref{eq:delta}] instead of $\ftopo$, since $\ftopo > 1$ when $S_{\mathrm{shuf}} < S_{\mathrm{rand}}$ (see text).}
\label{tab:ftopo}
\begin{ruledtabular}
\begin{tabular}{lcccccc}
Statistic & $L\!=\!16$ & $L\!=\!24$ & $L\!=\!32$ & $L\!=\!48$ & $L\!=\!64$ & $L\!=\!128$ \\
\hline
$\TP_{H_0}$ & 0.008 & 0.015 & 0.021 & 0.040 & 0.037 & 0.065 \\
$\TP_{H_1}$ & 0.044 & 0.071 & 0.093 & 0.123 & 0.133 & 0.186 \\
PE & 0.055 & 0.050 & 0.052 & 0.056 & 0.053 & 0.051 \\
$n_{H_0}$ & 0.000 & 0.000 & 0.000 & 0.000 & 0.000 & 0.000 \\
$n_{H_1}$ & 0.148 & 0.187 & 0.224 & 0.270 & 0.282 & 0.345 \\
$\delta(\max)$ & 0.77 & 1.77 & 2.58 & 3.50 & 4.67 & 8.07 \\
mean bar & 0.125 & 0.140 & 0.154 & 0.177 & 0.182 & 0.201 \\
\end{tabular}
\end{ruledtabular}
\end{table}

\subsection{The shuffled null detects $\Tc$}
\label{sec:published}

A central claim in the point-cloud PH literature is that PH detects $\Tc$~\cite{donato2016,sale2022}. We confirm this---but show that the shuffled null does too. Using finite differences on our ten-temperature grid ($\Delta T \approx 0.05$ near $\Tc$), the peak of $|d(\TP_{H_0}/L^2)/dT|$ falls in the same temperature bin for both real and shuffled configurations at $L = 32, 64$, and $128$ (Table~\ref{tab:peaks}). We cannot resolve sub-bin differences with this grid spacing, but the peak heights---a grid-independent quantity---agree to within 13\%. The mechanism is straightforward: since $\TP \approx g(\rho) L^2$ (see Sec.~\ref{sec:theory}), $d\TP/dT$ peaks where $d\rho/dT$ peaks---which is at $\Tc$. Point-cloud PH adds nothing beyond what $\rho(T)$ already provides for $H_0$ statistics.

\begin{table}[t]
\caption{Peak location and height of $|d(\TP_{H_0}/L^2)/dT|$ for real and shuffled configurations.}
\label{tab:peaks}
\begin{ruledtabular}
\begin{tabular}{lccccc}
$L$ & \multicolumn{2}{c}{Peak $T$} & \multicolumn{2}{c}{Peak height} & Ratio \\
& Real & Shuf & Real & Shuf & S/R \\
\hline
32 & 2.325 & 2.325 & 0.328 & 0.329 & 1.005 \\
64 & 2.284 & 2.284 & 0.544 & 0.614 & 1.128 \\
128 & 2.284 & 2.284 & 1.009 & 1.107 & 1.097 \\
\end{tabular}
\end{ruledtabular}
\end{table}

Notably, the shuffled peak is 10--13\% \textit{taller} than the real peak at $L = 64$ and $128$ (Table~\ref{tab:peaks}, ratio $> 1$). Spatial correlations in real configurations suppress $\TP_{H_0}$ relative to random placement at the same density---a consequence of clustering, which reduces the total number of short-range merges. This suppression was also observed in quench dynamics~\cite{loftus2026a}.

\begin{figure*}[t]
\includegraphics[width=\textwidth]{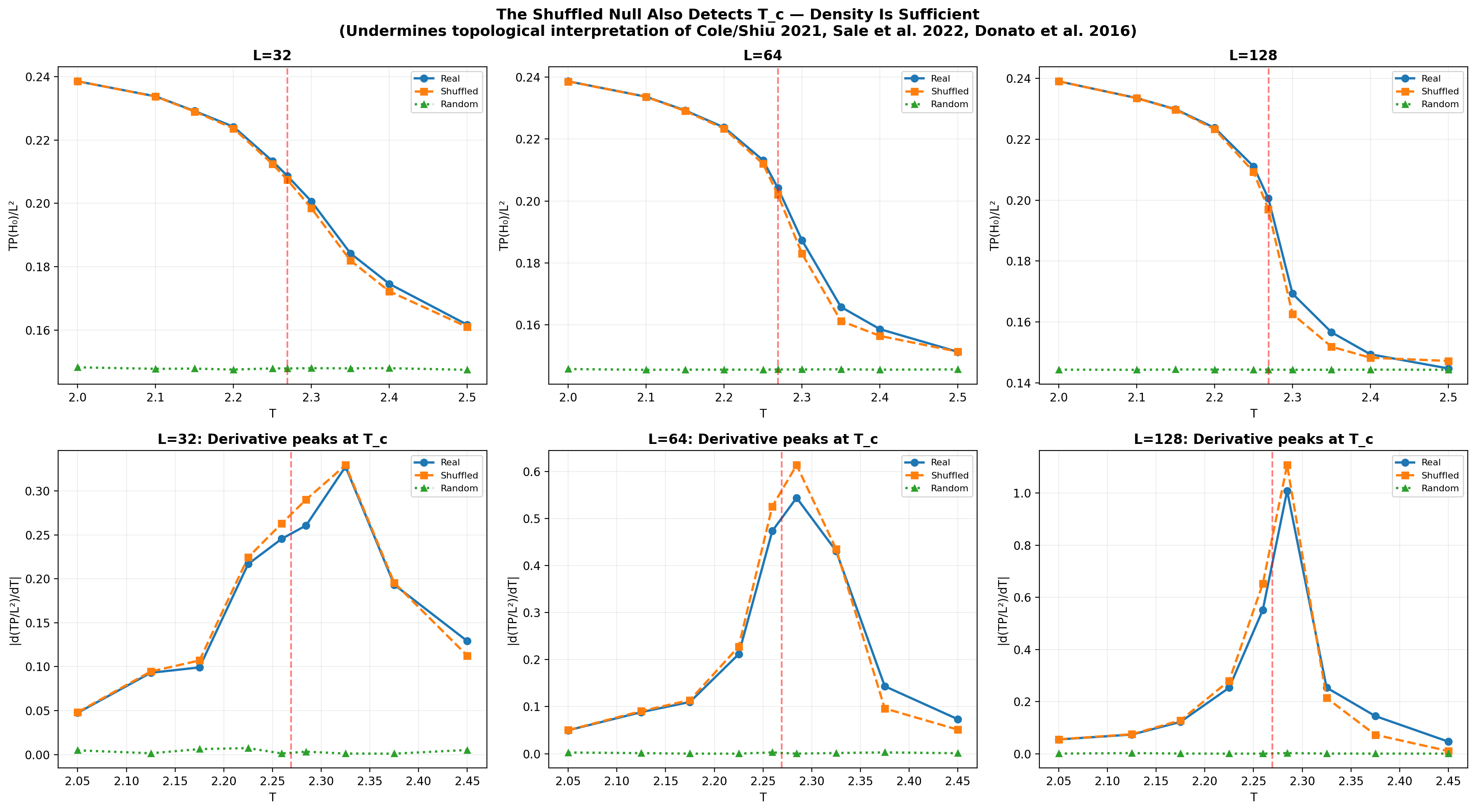}
\caption{The shuffled null detects $\Tc$. Top: $\TP_{H_0}/L^2$ vs.\ temperature for real (blue), shuffled (orange), and random (green) point clouds. Bottom: $|d(\TP_{H_0}/L^2)/dT|$ peaks at the same temperature bin for real and shuffled at all three system sizes. Peak heights agree to within 13\%. The random null (fixed $\rho = 0.5$) also peaks near $\Tc$ at $L = 128$.}
\label{fig:shuffled_tc}
\end{figure*}

This result challenges the topological interpretation of $\Tc$ detection in point-cloud PH studies. When such studies claim that ``PH detects the phase transition through topological features''~\cite{donato2016,sale2022}, the mechanism is more precisely described as: point-cloud PH detects the phase transition through \textit{density}, which is the physical order parameter (magnetization). The PD machinery adds nothing beyond what a simple measurement of the majority fraction $\rho(T)$ already provides for $H_0$ statistics. (We note that sublevel-set filtrations on the spin field~\cite{cole2021} are not subject to the same density confound, since all lattice sites are included at every temperature.)

\subsection{$H_1$ is partially topological and grows with $L$}
\label{sec:h1}

In contrast to $H_0$, the $H_1$ statistics show significant topological content that \textit{increases} with system size:

\begin{itemize}
\item $\ftopo(\TP_{H_1})$ grows from 0.044 ($L = 16$) to 0.186 ($L = 128$).
\item $\ftopo(n_{H_1})$ grows from 0.148 to 0.345.
\item The real-shuffled gap significance grows from 1.4$\sigma$ to 7.1$\sigma$ for $\TP_{H_1}$ and from 5.0$\sigma$ to 13.9$\sigma$ for $n_{H_1}$.
\end{itemize}

At $L = 64$, the alpha complex produces 2721 $H_1$ bars for real configurations vs.\ 2244 for shuffled---a 21\% excess of loop features arising from spatial correlations. This excess grows with $L$: the count ratio increases from 1.18 ($L = 32$) to 1.26 ($L = 128$).

The maximum bar length reveals the starkest contrast. Real configurations have $\max \approx 0.38 L$ (scaling with $L$), while shuffled configurations saturate at $\max \approx 5$ regardless of $L$. At $L = 128$, the real max bar is 8.8$\times$ larger than the shuffled max bar (48.5 vs.\ 5.5), a $> 26\sigma$ gap. This is a direct signature of the correlation length: real configurations at $\Tc$ have structure at scale $\xi \sim L$, producing long-lived $H_1$ features that random placement cannot generate.

\subsection{Finite-size scaling of the topological excess}
\label{sec:fss}

The relative topological excess [Eq.~\eqref{eq:delta}] at $\Tc$ follows a power law:
\begin{align}
\delta(\TP_{H_1}) &\sim L^{0.53 \pm 0.05} \quad (R^2 = 0.95), \label{eq:alpha_tp} \\
\delta(n_{H_1}) &\sim L^{0.41 \pm 0.03} \quad (R^2 = 0.97), \label{eq:alpha_n} \\
\delta(\max) &\sim L^{1.07 \pm 0.05} \quad (R^2 = 0.95). \label{eq:alpha_max}
\end{align}

The max-bar exponent $\alpha_{\max} \approx 1$ is expected: it reflects the correlation length $\xi \sim L^{1/\nu}$ with $\nu = 1$ for 2D Ising. The $\TP_{H_1}$ exponent $\alpha_{H_1} \approx 0.53$ is a new topological critical exponent. We caution that six data points provide limited discriminating power between functional forms: a logarithmic fit $\delta = a\ln L + b$ gives $R^2 = 0.996$ and a saturation model $\delta = 1 - cL^{-\beta}$ gives $R^2 = 0.995$, both slightly better than the power law. All three models predict $\delta \to \infty$ or $\delta \to 1$ as $L \to \infty$---i.e., topology dominates $H_1$ at large $L$---but the precise functional form requires data at larger system sizes.

\begin{figure}[t]
\includegraphics[width=\columnwidth]{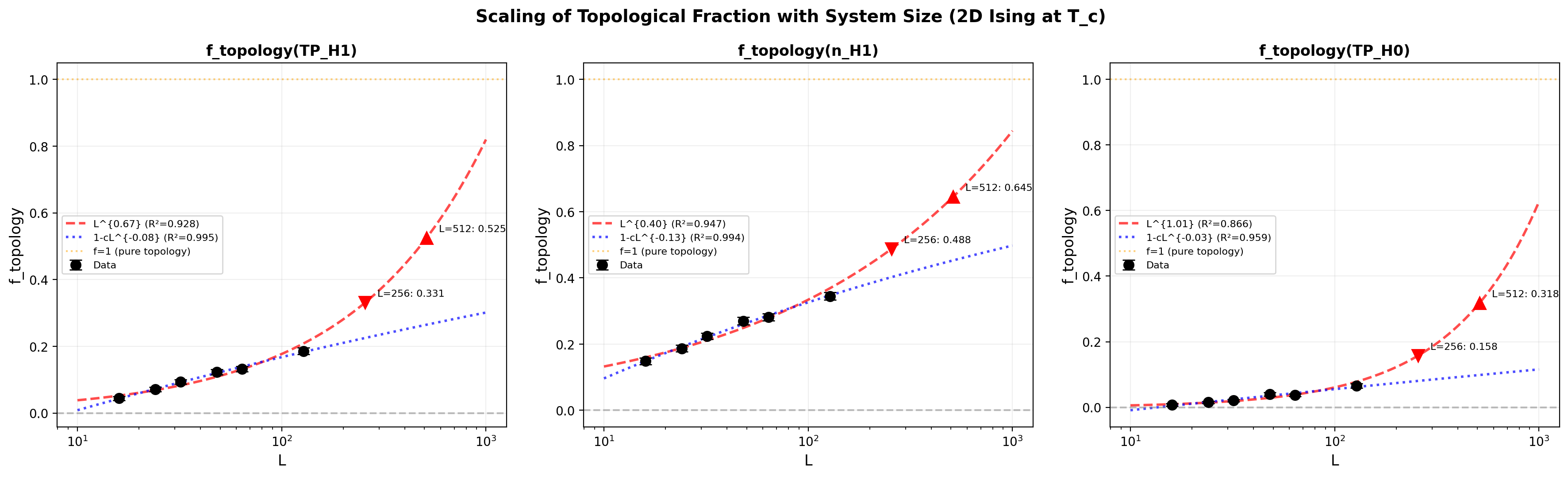}
\caption{Scaling of the topological fraction with system size at $\Tc$. Left: $\ftopo(\TP_{H_1})$ grows with $L$; a power law $L^{0.67}$ (red dashed) and saturation model (blue dotted) both fit the data and predict topology dominates $H_1$ at $L \to \infty$. Center: $\ftopo(n_{H_1})$ shows the same growth. Right: $\ftopo(\TP_{H_0})$ remains small at all $L$. Triangles mark extrapolated values at $L = 256$ and $512$.}
\label{fig:h1_scaling}
\end{figure}

To test whether $\delta$ follows a finite-size scaling (FSS) ansatz, we posit
\begin{equation}
\delta(T, L) = L^{\alpha} \, g\!\left(\frac{T - \Tc}{\Tc} \cdot L^{1/\nu}\right)
\label{eq:fss}
\end{equation}
with $\nu = 1$ and $\alpha$ from Eqs.~\eqref{eq:alpha_tp}--\eqref{eq:alpha_n}. Data from all six system sizes and ten temperatures collapse onto a universal scaling function $g$ with coefficient of variation CV $= 0.27$ for $\TP_{H_1}$ and CV $= 0.25$ for $n_{H_1}$. For comparison, typical FSS analyses of thermodynamic quantities achieve CV $\approx 0.1$--$0.2$, so our collapse is partial but meaningful.

\subsection{Theoretical foundation: universal density curve}
\label{sec:theory}

\begin{figure}[t]
\includegraphics[width=\columnwidth]{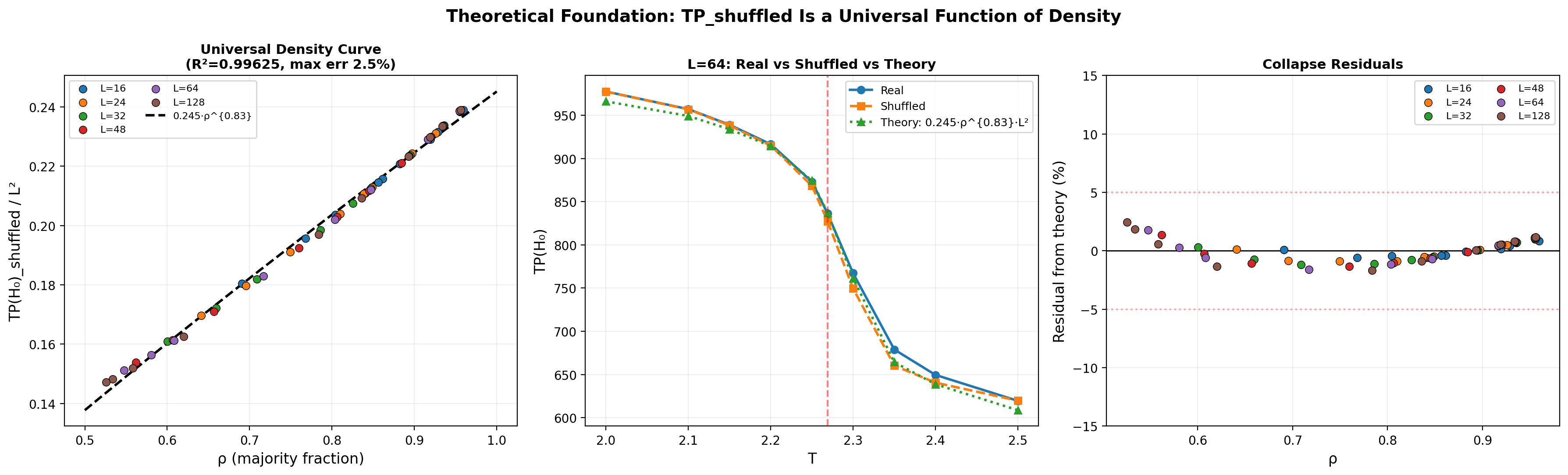}
\caption{Left: $\TP_{H_0}^{\mathrm{shuf}}/L^2$ vs.\ majority fraction $\rho$ for all 60 $(L, T)$ combinations (six system sizes, ten temperatures). All points collapse onto a single curve $0.245\,\rho^{0.83}$ (dashed, $R^2 = 0.996$). Center: real, shuffled, and theoretical prediction for $L = 64$. Right: residuals from the universal curve, all below 2.5\%.}
\label{fig:collapse}
\end{figure}

We show that the shuffled null's total persistence is a deterministic function of density. Define $\rho = n/L^2$ as the majority-spin fraction. Across all 60 $(L, T)$ combinations, the shuffled $\TP_{H_0}$ satisfies
\begin{equation}
\frac{\TP_{H_0}^{\mathrm{shuf}}}{L^2} = 0.245 \, \rho^{0.83} \quad (R^2 = 0.996).
\label{eq:collapse}
\end{equation}
The maximum residual is 2.5\%, and the collapse holds across all six system sizes (Fig.~\ref{fig:collapse}).

 This is consistent with known universality results for persistence of random point processes: Divol and Polonik~\cite{divol2019} proved that (suitably normalized) persistence functionals converge to deterministic limits for Poisson point clouds, and Skraba \textit{et al.}~\cite{skraba2024} showed that the limiting persistence measure is independent of the underlying density.

The practical implication is that $\TP^{\mathrm{shuf}}$ can be \textit{predicted} from $\rho$ alone, without computing any PD. The topological content of a real configuration is then the departure from this prediction:
\begin{equation}
\Delta \TP = \TP^{\mathrm{real}} - 0.245 \, \rho^{0.83} \, L^2.
\end{equation}

\subsection{Mechanism: scale-resolved topological excess}
\label{sec:mechanism}

Where in the persistence diagram does the topological excess reside? We bin $H_1$ bar lifetimes into logarithmic scale intervals and compute the TP contribution from each bin for real and shuffled configurations. The topological excess $\Delta\TP_{H_1}(\ell)$ in each bin reveals two mechanisms:

\textit{Long bars} ($\ell > L/10$): These are the correlation-length-scale loops. They contribute a decreasing fraction of the total excess as $L$ grows: 112\% at $L = 32$ (where small-scale excess is slightly negative) but only 40\% at $L = 128$.

\textit{Short/medium bars} ($\ell \leq L/10$): These are loops around individual minority-spin clusters. Their contribution grows with $L$: from $-12\%$ at $L = 32$ to 60\% at $L = 128$.

The median scale of the topological excess shifts: $\ell_{\mathrm{med}}/L = 0.20$ at $L = 32$, $0.13$ at $L = 64$, and $0.08$ at $L = 128$. This shift explains the FSS exponent: the excess TP grows both because more scales contribute (as $L$ increases, more decades of the cluster-size spectrum become available) and because more minority clusters of each size exist.

The shape of the $H_1$ lifetime distribution further clarifies the mechanism. Real and shuffled distributions have identical median lifetime ($\ell = 0.25$, the nearest-neighbor distance on the lattice) but differ dramatically in their tails. Real configurations produce bars extending to $\ell \approx 0.38 L$, while shuffled configurations are truncated at $\ell \approx 5$ regardless of $L$. The real distribution develops a power-law tail above $\ell \approx 1$, directly reflecting the fractal cluster structure of the critical Ising model.

\subsection{Potts models}
\label{sec:potts}

As a consistency check, we test whether the density-dominance pattern extends to other transition types. Using the corrected random baseline $\rho = 1/q$ for Potts models at a single system size $L = 48$ (insufficient for independent FSS analysis):

\textit{Potts $q = 3$} (second-order, $\Tc \approx 0.995$): At $T = 0.99$, $\ftopo(\TP_{H_0}) = 0.029$, confirming density dominance.

\textit{Potts $q = 5$} (first-order, $\Tc \approx 0.852$): At $T = 0.85$, $\ftopo(\TP_{H_0}) = 0.019$. The first-order transition, with its sharper density jump, is even more density-dominated.

Above $\Tc$, the disordered phase of Potts models has majority fraction $\rho \to 1/q$. Since the random baseline uses the same density, the $\ftopo$ decomposition is well-defined throughout.

\section{Discussion}
\label{sec:discussion}

\subsection{What PH detects and what it doesn't}

Our results establish a clear dichotomy. When PH is applied to spin model point clouds via the alpha complex:

$H_0$ statistics detect phase transitions through \textit{density}. The mechanism is indirect: the magnetization $m(T)$ determines the majority fraction $\rho(T)$, which determines the point count $n = \rho L^2$, which determines $\TP_{H_0} \approx g(\rho) L^2$. The PD is an expensive proxy for the order parameter.

$H_1$ statistics contain genuine topological information that grows with system size. Loop features in the alpha complex are sensitive to the spatial arrangement of points---specifically, to the minority-spin voids that create holes in the Delaunay triangulation. These voids have a fractal size distribution at $\Tc$, producing an $H_1$ topological excess that scales as $L^{0.53}$.

The maximum persistence bar is the strongest topological signal. It scales with $\xi$ and is 5--9$\times$ larger for real configurations than for shuffled, even at moderate $L$. This is consistent with the exp20b result~\cite{loftus2026a} that the maximum $H_0$ bar is 2.7$\times$ larger for real vs.\ shuffled across three lattice geometries.

\subsection{Relation to prior work}

Several authors have noted the importance of null models in TDA~\cite{bobrowski2017,divol2019,skraba2024}, but to our knowledge, no study of PH applied to spin model phase transitions has included a density-matched shuffled null. The closest is the work of Spitz \textit{et al.}~\cite{spitz2020}, who compared PH to non-topological descriptors, and Olsthoorn \textit{et al.}~\cite{olsthoorn2020}, who used machine learning on PDs without testing whether the signal is topological.

Our shuffled null is closely related to the Poisson null model in stochastic geometry~\cite{bobrowski2018}, with the additional constraint that points are placed on lattice sites (matching the geometric substrate of the real configurations).

\subsection{Scope and limitations}

Our analysis addresses point-cloud PH---the construction of alpha or Rips complexes on spin-position point clouds---which is the methodology of Donato \textit{et al.}~\cite{donato2016}, Sale \textit{et al.}~\cite{sale2022}, and others. A separate line of work uses sublevel-set filtrations on the spin field itself via cubical complexes~\cite{cole2021,olsthoorn2020}. In that setting, all $L^2$ lattice sites are included at every temperature and the filtration value is the spin variable, not a distance. The density confound identified here---varying point count as a function of temperature---does not apply to sublevel-set filtrations. Whether sublevel-set PH has its own confounds (e.g., sensitivity to the mean spin value rather than spatial correlations) is an open question that we do not address.

Additionally, our Potts results use a single system size ($L = 48$) and should be interpreted as consistency checks rather than independent demonstrations of universality. Finite-size scaling analysis for Potts models would require multiple system sizes.

\subsection{A new topological critical exponent}

The scaling $\delta(\TP_{H_1}) \sim L^{0.53}$ defines a topological critical exponent that, to our knowledge, has no known relation to the standard 2D Ising exponents ($\eta = 1/4$, $\nu = 1$, $\beta = 1/8$). We have attempted heuristic derivations via integration of pair correlation functions over scale, minority-cluster size distributions, and fractal boundary counts, but none reproduce the measured value. The alpha complex geometry---being the nerve of the Voronoi dual---introduces non-trivial dependence on local point configurations that resists simple analytical treatment.

We conjecture that $\alpha_{H_1}$ is a genuinely new exponent characterizing the topological content of alpha-complex PH at criticality. Its analytical derivation remains an open problem at the intersection of stochastic topology and critical phenomena.

\subsection{Recommendations for practitioners}

Based on our findings, we recommend:

\begin{enumerate}
\item \textbf{Always include a shuffled null.} Every claim that PH detects a phase transition should be tested against a density-matched null model. If the null also detects the transition, the signal is density-driven.
\item \textbf{Use $H_1$, not $H_0$.} Topological information resides primarily in loop features, not connected components. The $H_1$ topological fraction grows with system size, making it increasingly informative at larger $L$.
\item \textbf{Report $\ftopo$ alongside PH statistics.} The topological fraction quantifies what the PD contributes beyond density, enabling meaningful cross-study comparisons.
\item \textbf{Focus on the maximum bar.} It is the strongest topological signal, scales with $\xi$, and is easily interpretable.
\end{enumerate}

\section{Conclusion}
\label{sec:conclusion}

We have shown that persistent homology applied to spin model point clouds is 94--100\% density-driven for $H_0$ statistics and partially topological for $H_1$ statistics, with the topological fraction growing as a power law in system size. The shuffled null detects $\Tc$ at the same location as real configurations, demonstrating that the widely celebrated ``topological detection'' of phase transitions is, for standard $H_0$-based analysis, a density effect.

The constructive message is not that PH is useless---it is that PH works for a different reason than previously believed. Once the density contribution is subtracted, the topological residual reveals genuine spatial structure: loop features arising from minority-spin cluster boundaries, scaling with the correlation length, and following finite-size scaling with a new critical exponent $\alpha_{H_1} \approx 0.53$. Extracting this residual requires the shuffled null that we advocate as standard practice.

\begin{acknowledgments}
Computations used GUDHI~\cite{gudhi2015} for persistence diagrams and the Wolff cluster algorithm~\cite{wolff1989} for Monte Carlo sampling.
\end{acknowledgments}

\bibliography{density_topology}

\end{document}